\documentclass{aa}  

\usepackage{graphicx}
\usepackage{float}
\usepackage{txfonts}
\usepackage[hidelinks]{hyperref}
\usepackage{siunitx}

\begin{document}

\title{Constraining the presence of exotrojans in hot Jupiter systems using TTV observations from TESS}

\author{Zixin Zhang
\inst{1,2}
\and
Wenqin Wang
\inst{1,2}
\and
Xinyue Ma
\inst{3,4}
\and
Zhangliang Chen
\inst{1,2}
\and
Yonghao Wang
\inst{5}
\and
Cong Yu
\inst{1,2}
\and
Shangfei Liu
\inst{1,2}
\and
Yang Gao
\inst{1,2}
\and
Baitian Tang
\inst{1,2}
\and
Dichang Chen
\inst{1,2}
\and
Bo Ma
\inst{1,2}
\fnmsep\thanks{Corresponding author: {mabo8@mail.sysu.edu.cn}}
}

\institute{School of Physics and Astronomy, Sun Yat-sen University,
  Zhuhai 519082, China \\
  \and
  CSST Science Center for the Guangdong-Hong Kong-Macau Great Bay Area,
  Sun Yat-sen University, Zhuhai 519082, China
  \and
  Instituto de Astrofísica de Canarias (IAC), Vía Láctea s/n, 38205 La Laguna, Tenerife, Spain
  \and
  Departamento de Astrofísica, Universidad de La Laguna (ULL), C/ Padre Herrera, 38206 La Laguna, Tenerife, Spain
  \and
  Hainan University, China
}

\date{Accepted for publication in Astronomy \& Astrophysics}
\idline{Astron. Astrophys., accepted (aa57704-25)}

\abstract
{Co-orbital objects, also known as Trojans, are celestial bodies that share a 1:1 mean motion resonance orbit with a planet. Many planets in our solar system hold Trojans, yet there have been no confirmed ones in exoplanetary systems. While theoretical models suggest that hot Jupiters are unlikely to retain co-orbitals, their high transit cadence and deep transits make them accessible targets for constraining the existence of massive companions using Transit Timing Variations (TTVs).}
{The ExoEcho (Exoplanet Ephemerides Change Observations) project aims to study TTVs in exoplanetary systems using high-precision photometry from space and ground-based telescopes. As the third paper in the ExoEcho project, we investigate the potential existence of exotrojans in hot Jupiter systems through TTV analysis of TESS observations.}
{In this work, we analyze TESS photometry for 260 confirmed hot Jupiters with published RV-based mass measurements to search for TTV signals compatible with Trojan companions. We derived transit times and assessed the compatibility of TTV residuals with co-orbital models through N-body simulation using the REBOUND N-body code. Accounting for the physical degeneracy between the Trojan mass and the libration amplitude, we place upper mass limits on possible companions for typical libration amplitudes. }
{For a typical libration amplitude of $15^\circ$, we rule out exotrojans more massive than 1 Earth mass in 130 systems ($\sim50\%$ of our sample). A more conservative $\chi^2$ analysis incorporating observational uncertainties places this limit at 3 Earth masses. Additionally, we use the stability boundaries of the 1:1 resonance to exclude dynamically unstable configurations, ensuring our derived mass limits fall within the stable regime.
}
{Our results provide stringent constraints on co-orbital formation in short-period systems and establish a framework for future high-precision missions such as PLATO (PLAnetary Transits and Oscillations of stars) or ET (Earth 2.0) mission.}


\maketitle

\section{Introduction}

Co-orbital objects, also known as Trojans, are celestial bodies that share a 1:1 mean-motion resonance with a planet. They typically reside near the system's Lagrange points \(L_4\) and \(L_5\), approximately \(60^\circ\) ahead of and behind the planet along its orbit. Depending on their orbital behavior relative to the planet, co-orbital bodies can follow two main types of trajectories: tadpole orbits and horseshoe orbits \citep{garfinkel1976-theory, erdi1977-asymptotic, garfinkel1978-theory, niederman2020-coorbital, raymond2023-constellations}. In a tadpole orbit, the object librates around either the \(L_4\) or \(L_5\) point with a small angular amplitude, forming a path that resembles a tadpole shape in a rotating reference frame. In contrast, objects on horseshoe orbits exhibit much larger oscillations in angular separation, traveling along a broad arc that encompasses the \(L_3\), \(L_4\), and \(L_5\) points, effectively tracing a horseshoe-shaped path relative to the planet. Tadpole orbits are typically more robust than horseshoe orbits, because the horns of the horseshoe approach too close to the planet (in terms of the planet’s Hill radii) for more massive planets \citep{laughlin2002-extrasolar, cuk2012-longterm}. Early Trojan progenitors on horseshoe orbits can be captured into stable tadpole configurations during Jupiter's gas accretion and mass growth \citep{fleming2000-origin, raymond2023-constellations}.

The existence of Trojan planets in exoplanetary systems, known as exotrojans, has significant implications for our understanding of planet formation and migration processes \citep{chiang2005-neptune}. Similar to their counterparts in our solar system, exotrojans are thought to be by-products of early planetary evolution. Thus, studying these celestial bodies can offer valuable insights into the initial stages of planetary system development \citep{lillo-box2018-troy}. Several theories have been proposed to explain the formation of exotrojans.
Two prominent scenarios exist: the in situ formation scenario, where exotrojans are remnants of early planet formation \citep{laughlin2002-extrasolar, chiang2005-neptune, lykawka2010-capture}, and the late-capture scenario, where they are captured during planetary migration \citep{laughlin2002-extrasolar,chiang2005-neptune,beauge2007-coorbital,lyra2009-standing}.

Various techniques have been developed to detect exotrojans. One prominent method, introduced by \citet{ford2006-observational}, combines transit and radial velocity (RV) observations to search for exotrojans and constrain the mass of a possible co-orbital body. Another powerful approach involves analyzing transit timing variations (TTVs), which are deviations from a planet's strictly periodic transit schedule. The gravitational pull from a co-orbital companion perturbs the transiting planet's orbit, causing its transits to occur slightly earlier or later than predicted. \citet{ford2007-using} first explored this effect for exotrojans, demonstrating that even a sub-Earth-mass companion could produce a TTV signal detectable by ground-based observatories. Subsequent studies, such as \citet{haghighipour2013-detection}, reinforced this by showing that an Earth-mass exotrojan could induce TTVs with amplitudes of up to a few hours, well within the sensitivity of current ground-based telescopes.

Previous systematic searches for Trojans in the Kepler data have been conducted by \citet{janson2013-systematic}. To extend the search for co-orbital companions such as exotrojans and exomoons, \citet{leleu2017-detection} introduced the $\alpha$-test method. More recently, the RIVERS project \citep{leleu2021-rivers, leleu2022-alleviating} has further advanced the detection of co-orbitals using TTVs. Building on these methods, the TROY Project \citep{lillo-box2018-troy, lillo-box2018-troya, balsalobre-ruza2024-project} systematically searches for exotrojan companions using a combination of transit, RV, and TTV techniques.

The TROY Project has systematically explored the population of Trojan bodies in exoplanetary systems. In its latest study, \citet{balsalobre-ruza2024-project} analyzed 95 transiting planets across 84 systems (focusing on low-mass stars) using the $\alpha$-test method, setting upper mass limits for potential co-orbital companions and ruling out the presence of exotrojans heavier than Saturn. A strong candidate was identified in the GJ 3470 system, which marks the potential first detection of an exotrojan companion in an exoplanetary system. Furthermore, recent advancements in direct imaging have enabled tentatively probing co-orbital submillimeter emission in young systems \citep{2023A&A...675A.172B, 2022ApJ...937L...1L}. These efforts demonstrate the growing capabilities in detecting and characterizing exotrojan companions.

The ExoEcho (Exoplanet Ephemerides Change Observation) project aims to study exoplanet photodynamics using precise transit timing data from ground- and space-based telescopes \citep{wang2023-longterm, zhang2024-constraining, Ma24, wang_exoplanet_2025}. As part of the ExoEcho project, this study investigates the possible existence of Trojan companions in hot Jupiter systems using TTV observations from the Transiting Exoplanet Survey Satellite \citep[TESS;][]{ricker2014-transiting}. TESS, launched in 2018 as a successor to the Kepler space telescope \citep{borucki2010}, has significantly advanced exoplanet research with its high-precision, all-sky photometric survey. Designed to expand our understanding of exoplanets through the transit method, TESS surveys nearly the entire sky, including the Continuous Viewing Zones at the ecliptic poles. It yields a photometric precision of approximately 700 ppm for a 12th-magnitude star in a 1-hour integration \citep{ricker2014-transiting}.
These capabilities make TESS an ideal instrument for detecting exoplanets using transit techniques and exploring the broader landscape of exoplanetary science through photodynamics techniques. While hot Jupiters are not dynamically favorable for retaining primordial Trojans \citep{leleu2019-stability}, they serve as accessible targets for observational searches due to their high transit cadence.

This paper is structured as follows. Sect.~\ref{sec:data} presents the TESS photometric observations of 260 hot Jupiter systems used in our analysis. Sect.~\ref{sec:result} details the method for deriving upper mass limits for potential exotrojans and presents the results. We discuss the implications of these findings in Sect.~\ref{sec:discussion}, and Sect.~\ref{sec:conclusion} summarizes the main conclusions of this study.

\section{Data and observation}
\label{sec:data}
\begin{figure}
  \centering
  \includegraphics[width=0.45\textwidth]{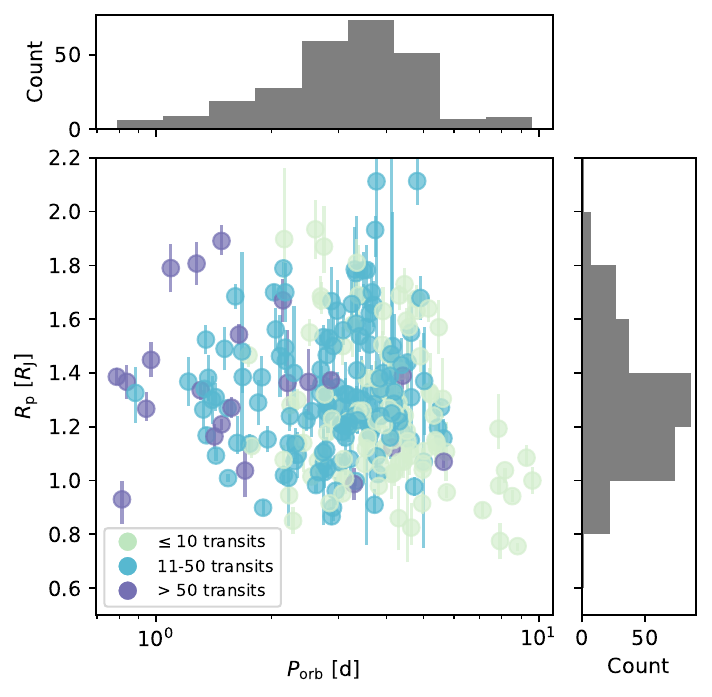}
  \caption{Radius and period properties of the sample of 260 hot-Jupiter systems studied in this work. Most of them have orbital periods ranging from 2 to 6 days and radii between 0.8 and 1.6~$R_{\mathrm{J}}$. The numbers of transits are shown in different colors.}
  \label{fig:orbital-parameters}
\end{figure}

The data used in this study were obtained from observations by TESS. We selected confirmed transiting hot Jupiters with published mass determinations based on the following criteria: an orbital period shorter than 10 days and a planetary mass exceeding $0.25~M_{\mathrm{J}}$. To ensure a robust sample, only systems with at least three observed transits are included. This selection yields a sample of 260 confirmed hot Jupiter systems with RV-based mass measurements, representing one of the largest datasets used in similar studies to date.
In Fig.~\ref{fig:orbital-parameters}, we show the distribution of the planetary orbital parameters, indicating that most systems in our sample have orbital periods between 2 and 6 days, with planetary radii ranging from 0.8 to 1.6~$R_{\mathrm{J}}$.

\subsection{Data acquisition}
We use Astroquery \citep{ginsburg2017-astroquery} to access TESS light curve products for hot Jupiters produced by the Science Processing Operations Center \citep[SPOC;][]{Jenkins16-spoc} from the Mikulski Archive for Space Telescopes \citep[MAST;][]{BrasseurDonaldson2019, mast_team_2021}. The analysis involves using pre-search data conditioning simple aperture photometry (PDCSAP) light curves from SPOC \citep{2012PASP..124..985S, 2012PASP..124.1000S, 2014PASP..126..100S, Jenkins16-spoc} with a 2-minute cadence, where common instrumental systematic errors have been removed.
Figure~\ref{fig:orbital-parameters} also shows the distribution of the number of transits in the TESS observations data used for our sample of 260 hot Jupiter systems. Most systems have fewer than 100 transits, while some exceed this count. In particular, Qatar-10 b stands out with more than 200 transits due to its short period (1.6 days) and its specific location in the sky \citep{alsubai2019-qatar}.

\subsection{Transit modeling and TTV extraction}
We began by fitting the stitched light curve from all sectors of each hot Jupiter with the \texttt{PyTransit} package \citep{parviainen2015-pytransit}, which employs a Markov Chain Monte Carlo (MCMC) algorithm, to obtain preliminary estimates of the transit ephemerides.

We fitted a \citet{mandel2002-analytic} model to the phase-folded light curve of each TESS sector. The parameters used for the transit model included the time of inferior conjunction $T_0$, orbital period $P$, stellar density $\rho_\star$ (in $\si{g\,cm^{-3}}$), impact parameter $b$, and planet-to-star area ratio $(R_{\mathrm{p}}/R_\star)^2$. We assigned normal priors to $T_0$ and $P$, and uniform priors to $\rho_\star$, $b$, and $(R_{\mathrm{p}}/R_\star)^2$. The eccentricity and argument of periastron were fitted using the parameterization of $e \cos w$ and $e \sin w$. The quadratic limb-darkening coefficients were fitted using the $q_1$ and $q_2$ parameterization \citep{kipping2013-efficient} with uniform priors.

Most MCMC priors were derived from online catalog parameters (e.g., the means and standard deviations of the normal priors for period and $T_0$ match the values from the respective discovery papers). The prior for the mid-transit time of each event was centered on a generic linear ephemeris derived from literature values, allowing each transit to be modeled individually. We used the \texttt{PyTransit} implementation of the differential evolution global optimization algorithm followed by an MCMC sampling. We ran 100 chains with 2000 steps each. After the MCMC analysis, we checked the corner plot to ensure that the chains had converged.

\begin{figure*}
  \sidecaption
  \includegraphics[width=0.68\textwidth]{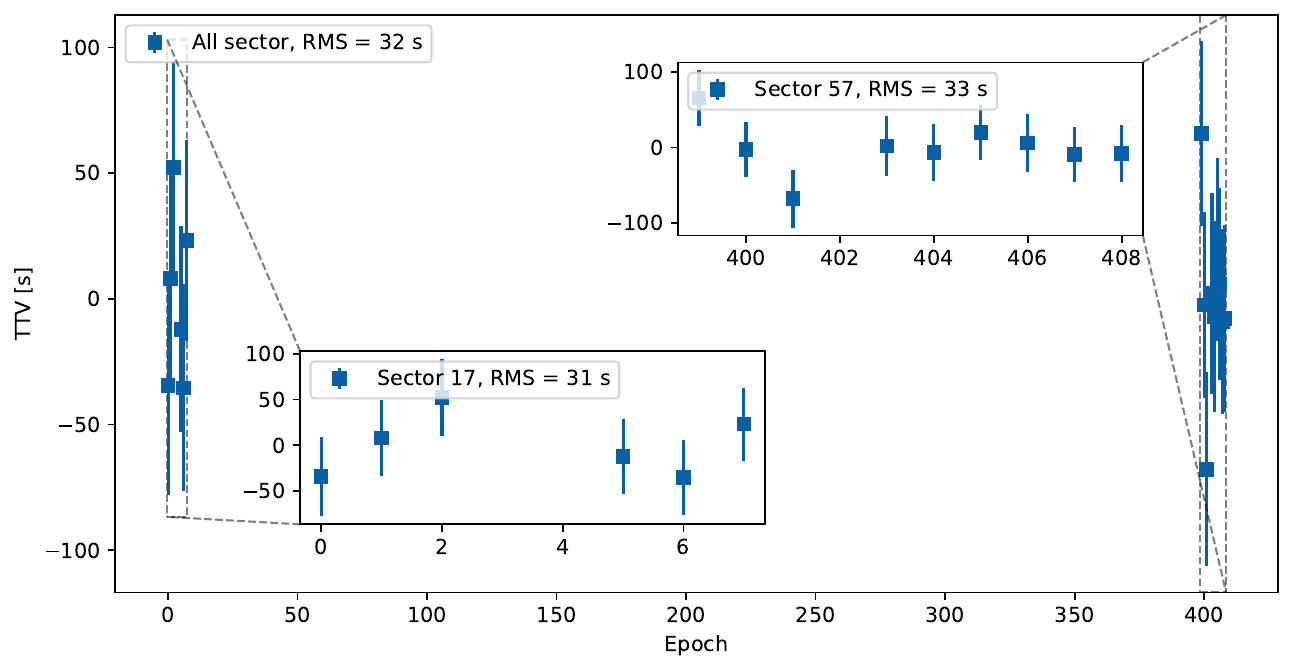}
  \caption{TTV measurements for WASP-93~b derived using a linear ephemeris model. The blue square represents the combined TTV obtained from TESS sectors 17 and 57. The insets show individual TTV measurements from each observation within these sectors, offering a zoomed-in view. The overall RMS for the TTVs is 32 seconds, with individual sectors showing an RMS of 31-32 seconds. Epoch 0 corresponds to the reference time of inferior conjunction at BJD 2458765.64897.}
  \label{fig:WASP-93 b-ttv}
\end{figure*}

Next, to precisely measure the mid-transit time for each individual transit, we segmented the TESS light curve into smaller windows, each spanning three times the transit duration and centered on the expected transit time. We then fitted the light curve within each window using \texttt{PdotQuest} \citep{wang2023-longterm}.
The transit timing variations (TTVs) are calculated by subtracting a linear ephemeris from the observed mid-transit times of each hot Jupiter. As an example, Fig.~\ref{fig:WASP-93 b-ttv} shows the TTVs for WASP-93 b, derived from a linear model. The measured mid-transit times and TTVs for WASP-93 b are listed in Table \ref{tab:ttv-data} as an example for the machine-readable table available at the CDS.

\begin{table*}
  \centering
  \caption{Transit times and TTVs for WASP-93 b.}
  \label{tab:ttv-data}
  \begin{tabular}{cccc}
    \hline\hline
    Planet    & Epoch & $T_c$ [BJD]   & $\sigma_{T_c}$ [days] \\
    \hline
    WASP-93 b & 0     & 2458765.64897 & 0.00050               \\
    WASP-93 b & 1     & 2458768.38200 & 0.00048               \\
    WASP-93 b & 2     & 2458771.11505 & 0.00049               \\
    WASP-93 b & 5     & 2458779.31192 & 0.00047               \\
    WASP-93 b & 6     & 2458782.04419 & 0.00048               \\
    ...       & ...   & ...           & ...                   \\
    \hline
  \end{tabular}
  \tablefoot{The full table is available at the CDS.}
\end{table*}

\section{Analysis and results}
\label{sec:result}

\subsection{TTV methodology}
\begin{figure}
  \centering
  \includegraphics[width=0.45\textwidth]{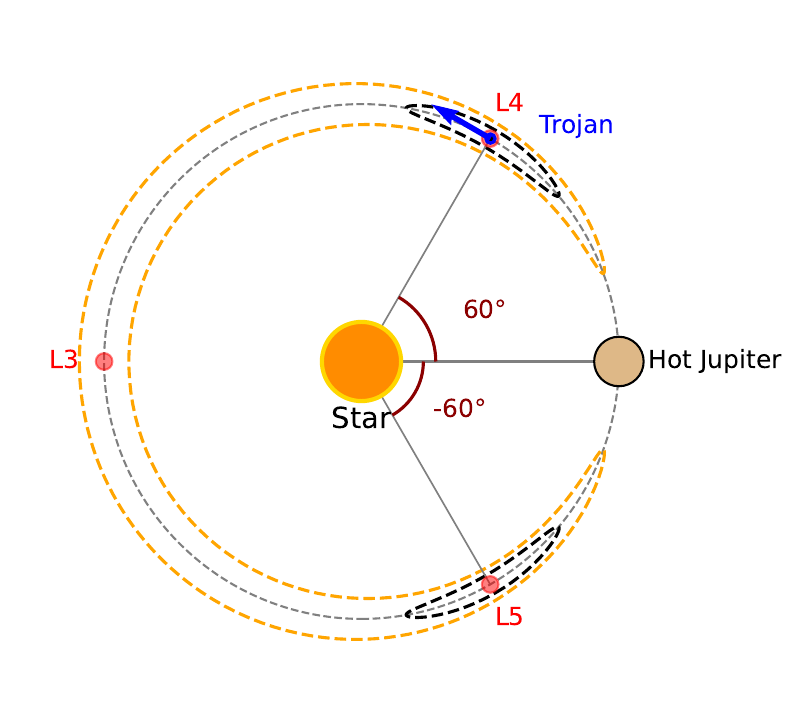}
  \caption{Schematic illustration of exotrojan orbits near a hot Jupiter. The figure shows both tadpole orbits, which librate around the Lagrangian points $L_4$ and $L_5$, and horseshoe orbits, which encompass a horseshoe-shaped region from $L_4$ to $L_5$. The host star and hot Jupiter are indicated, and the rotating reference frame is centered on the star-hot Jupiter barycenter.
  }
  \label{fig:trojan-illustrate}
\end{figure}

As demonstrated by \citet{ford2007-using}, dynamical perturbations from an exotrojan librating around the $L_4/L_5$ Lagrange points (as illustrated in Fig.~\ref{fig:trojan-illustrate}) will cause deviations in the strictly periodic transit times of the primary planet.
These deviations manifest as a TTV signal with an approximate amplitude of:
\begin{align}
  A_{\mathrm{TTV}}\simeq60\,\mathrm{s}\,\left(\frac{P}{4~\mathrm{d}}\right)\left(\frac{M_{\mathrm{Troj}}}{M_\oplus}\right)\left(\frac{0.5~M_{\mathrm{J}}}{M_{\mathrm{p}}+M_{\mathrm{Troj}}}\right)\left(\frac{K_{\Delta M}}{10{\si{\degree}}}\right),
  \label{eq:ttv-amplitude}
\end{align}
where $M_{\mathrm{p}}$ and $M_{\mathrm{Troj}}$ denote the masses of the primary planet and the exotrojan companion, respectively; $P$ is the orbital period of the primary; and $K_{\Delta M}$ represents the libration amplitude of the exotrojan.
This method has inherent limitations, most notably that the TTV signal is directly coupled to the companion's libration amplitude.
Consequently, exotrojans with vanishingly small libration amplitudes—those residing at or very close to the triangular Lagrange points—produce negligible TTV signals and remain undetectable regardless of their mass.
Furthermore, Eq.~\ref{eq:ttv-amplitude} reveals a physical degeneracy between the Trojan mass and libration amplitude, implying that our derived mass limits depend on the assumed libration.

Due to the symmetry of the restricted three-body problem in the circular approximation, Eq.~\ref{eq:ttv-amplitude} applies to both $L_4$ and $L_5$ configurations.
Although the phase of the TTV signal differs between a leading ($L_4$) and a trailing ($L_5$) companion, our methodology relies on the RMS of TTV residuals to establish upper mass limits.
This approach is therefore insensitive to the specific location ($L_4$ versus $L_5$) and constrains the presence of a co-orbital companion at either point capable of inducing TTVs above the noise floor.
We also note that a symmetric population of Trojans at both $L_4$ and $L_5$ would dampen the net TTV signal, constituting another limitation of this technique.

We also consider the impact of calculating planet masses from radial velocity (RV) observations.
In systems containing an unaccounted massive Trojan, the observed RV amplitude reflects the combined mass of the planet and the Trojan ($M_{cat} \approx M_{\mathrm{p}} + M_{\mathrm{Troj}}$).
As shown in Eq.~\ref{eq:ttv-amplitude}, the TTV amplitude scales with the mass ratio $M_{\mathrm{Troj}} / (M_{\mathrm{p}} + M_{\mathrm{Troj}})$.
Since the catalog mass $M_{cat}$ overestimates the true planet mass $M_{\mathrm{p}}$, using it in the denominator leads to a reduction in the predicted TTV amplitude for a given Trojan mass.
Consequently, a larger Trojan mass would be necessary to produce a detectable signal.
Thus, our derived upper mass limits, which rely on these catalog masses, are actually conservative values for this upper limit.

\subsection{N-body simulations}
We then performed N-body simulations using \texttt{REBOUND} \citep{rein2012-rebound} to model a system that features an exotrojan companion and to assess its ability to reproduce the observed data. The system's initial conditions were derived from the best-fit parameters of the primary planet, assuming circular orbits for both the hot Jupiter and the Trojan. The exotrojan companion was placed near the 1:1 mean motion resonance (MMR) with the primary planet at the $L_4$ point ($60^\circ$ leading the planet). A further discussion of the impact of the circular orbit assumption is detailed in Sect.~\ref{sec:discussion}. To explore a range of libration amplitudes, we varied the initial osculating period ratio between the companion and the planet $P_{\mathrm{T}}/P_{\mathrm{p}}$ from 0.9 to 1.1 in steps of 0.005. This initial period-ratio scan is an indirect way to sample the resulting libration amplitudes of near-co-orbital configurations, rather than a direct specification of a physical period ratio away from the 1:1 resonance. Some initial conditions in this scan correspond to chaotic, horseshoe-like, or unstable trajectories and are therefore excluded by the stability filtering described below. For each configuration, the libration amplitude $K_{\Delta M}$ was determined as the RMS of the relative mean longitude deviation from the Lagrangian point ($\sigma_{\lambda_{\mathrm{troj}} - \lambda_{\mathrm{planet}}}$). We then calculated the corresponding TTV signals and derived mass constraints for each configuration.

\begin{figure*}
  \sidecaption
  \includegraphics[width=0.68\textwidth]{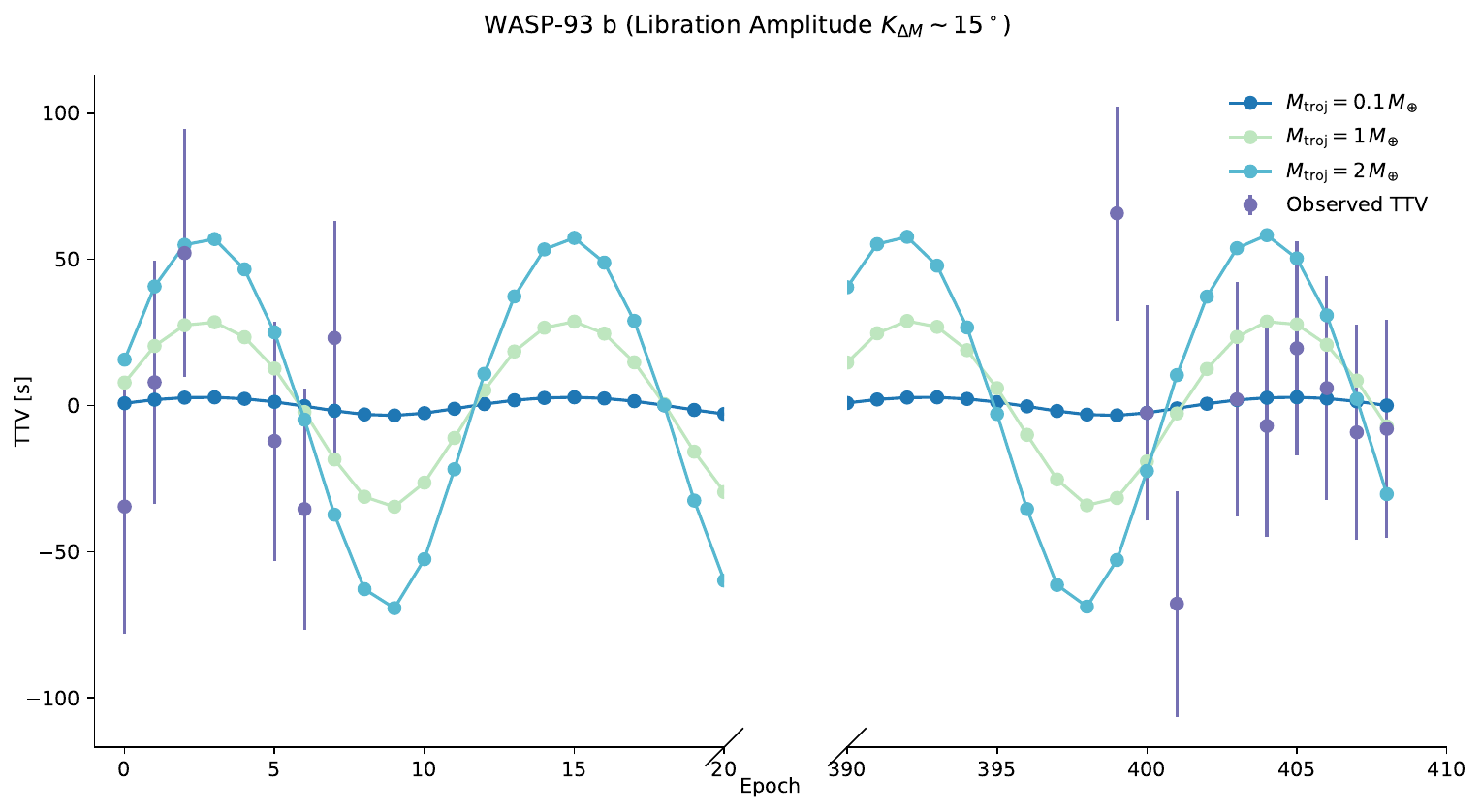}
  \caption{
    Observed TTV of the hot Jupiter system WASP-93~b, compared with simulated TTV signals induced by co-orbital exotrojans of varying masses. Different curves represent exotrojan masses ranging from Earth-mass to sub-Neptune mass, illustrating the mass-dependent amplitude and periodicity of the induced TTV signals.
  }
  \label{fig:wasp-93 b-ttv-rebound}
\end{figure*}

We note that generating initial conditions based on the exact libration amplitude would be a more direct approach. For instance, varying the initial angular separation $\zeta$ from the Lagrange point (as adopted in e.g., \citealt{leleu2019-stability}) is equally straightforward to set up in \texttt{REBOUND}. Other alternatives, such as perturbing the initial velocity, also exist. We have verified that all three approaches---varying the period ratio, the angular separation, or the velocity---produce equivalent TTV signals for a given libration amplitude. None of these methods guarantees an exact match to a desired libration amplitude due to the chaotic nature of $N$-body interactions, and our systematic scan of initial period ratios effectively covers the relevant range of libration amplitudes. A drawback of our simplified method is that some initial conditions with relatively large period deviations may correspond to chaotic or unstable orbits. To address this, we integrated a subsequent stability selection step (detailed in Sect.~\ref{sec:stability}) to exclude any inherently unstable configurations from our final mass limit statistics.
The duration of the simulated data was set to twice that of the TESS observations to avoid underestimating the RMS amplitudes. Fig.~\ref{fig:wasp-93 b-ttv-rebound} shows the simulated TTV amplitudes as a function of the exotrojan companion mass, compared to the observed data.

The RMS method has been widely used to analyze the TTV observation data of hot Jupiters in the past. For example, HAT-P-25~b is a hot Jupiter discovered by \citet{quinn2011-hatp25b}. \citet{wang_transiting_2018} have utilized the RMS approach to establish the maximum mass limit of HAT-P-25~b's companion and have ruled out any hidden long-period perturber with $M_{\mathrm{pert}} > 0.5, 0.3,$ and $0.5 M_{\mathrm{\oplus}}$ near the 1:2, 2:1, and 3:1 resonances.

\begin{figure}
  \centering
  \includegraphics[width=0.45\textwidth]{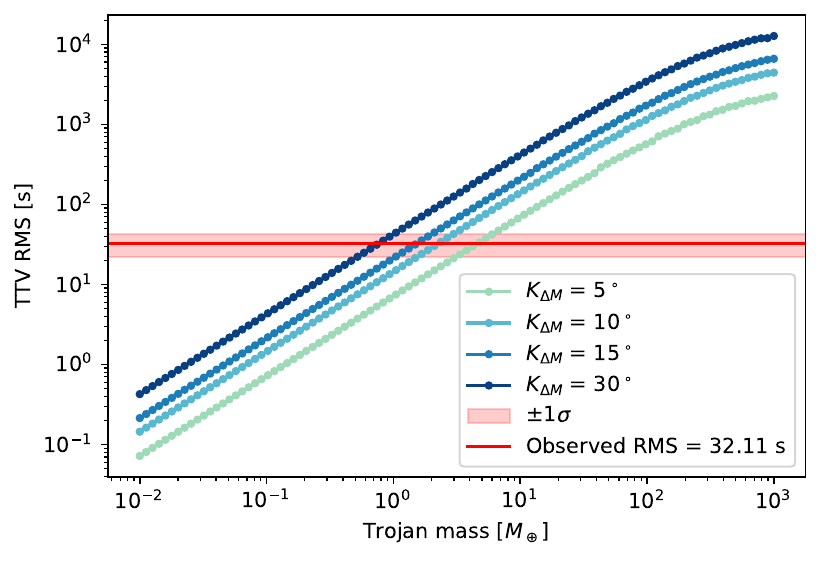}
  \caption{RMS variations of simulated TTVs as a function of exotrojan mass for WASP-93 b, assuming different libration amplitudes. The colored curves represent different libration amplitudes ($K_{\Delta M}$). The horizontal red dashed line indicates the observed TTV RMS. The intersections determine the upper mass limits for each specific libration amplitude. The $1\sigma$ uncertainty of the observed RMS, calculated via error propagation, determines the lower and upper bounds of the derived mass upper limit.
  }
  \label{fig:wasp-93 b_ttv_rms}
\end{figure}

Figure \ref{fig:wasp-93 b_ttv_rms} illustrates the RMS variations of simulated TTVs as a function of exotrojan mass for different libration amplitudes ($K_{\Delta M}$). The intersections of these curves with the observed TTV RMS level (horizontal dashed line) define the upper mass limits corresponding to each libration amplitude. This demonstrates the degeneracy between the companion's mass and its libration amplitude: larger libration amplitudes require smaller masses to produce the same TTV signal.
This mass sampling strategy aligns with RMS-based sensitivity analyses for companion mass constraints in previous TTV studies \citep{wang_transiting_2018, wang2021-transiting}. The above procedure was performed using our newly developed tool CMAT \citep[Companion MAss from Ttv modeling,][]{zhang_2024_13906089}\footnote{Available at \url{https://github.com/troyzx/CMAT}}, a Python package designed to automate the comparison between N-body simulations and observational data to constrain the mass of planetary companions (exotrojans in this case).
The light-curve modeling and dynamic simulation in this work closely follow the methods used in our previous paper \citep{zhang2024-constraining}.

To explicitly account for the uncertainties in the TTV measurements, we propagate the individual TTV errors to the final RMS calculation using standard uncertainty propagation. The uncertainty in the observed RMS ($\sigma_{\mathrm{RMS}}$) is calculated as:
\begin{equation}
  \sigma_{\mathrm{RMS}} \approx \frac{1}{N \cdot \mathrm{RMS_{obs}}}\left[\sum_{i=1}^{N} (\mathrm{TTV}_i\cdot \sigma_i)^2\right]^{1/2},
\end{equation}
where $\mathrm{TTV}_i$ and $\sigma_i$ are the individual TTV measurements and their uncertainties, and $N$ is the number of data points. This uncertainty is then projected onto the mass-libration plane to determine the lower and upper bounds of the derived upper mass limit at the $1\sigma$ level. Additionally, we performed a $\chi^2$ test to complement the RMS method. For each point in the parameter space (exotrojan mass vs. libration amplitude), we calculated the reduced $\chi^2$ statistic ($\chi^2_{\nu}$) comparing the observed TTVs with the simulated ones. This resulted in a $\chi^2$ map with libration amplitude on the x-axis and exotrojan mass on the y-axis, where the color contours indicate the goodness of fit. This approach allows us to determine the compatibility limits with statistical confidence, incorporating the observational errors.

\subsection{Stability analysis}
\label{sec:stability}

The stability of a system with an equilateral configuration, comprising a star and two planets of equal mass, was investigated by \citet{gascheau1843}.
The system is linearly stable for mass ratios
\begin{equation}
  \frac{M_{\mathrm{p}} + M_{\mathrm{Troj}}}{M_{\star} + M_{\mathrm{p}} + M_{\mathrm{Troj}}} < 0.038,
  \label{eq:stability}
\end{equation}
where $M_{\mathrm{p}}, M_{\mathrm{Troj}}$ are the masses of the primary planet and the exotrojan companion, respectively, and $M_{\star}$ is the mass of the host star. We used this criterion to ensure that all the orbital configurations studied are stable.

In addition to the analytical criterion, we performed numerical stability analysis using the Mean Exponential Growth factor of Nearby Orbits (MEGNO) indicator \citep{cincotta2003-phase}. For each configuration, we integrated the system for 10,000 orbits and calculated the MEGNO value. We excluded any configuration with a MEGNO value greater than 3, which indicates a chaotic or unstable orbit. While integrations spanning $10^4$ orbits (typically hundreds of years for short-period planets) are short compared to the system's lifetime, for these close-in hot Jupiters, the $L_4/L_5$ libration periods are typically only a few dozen planetary orbits. This means $10^4$ planetary orbits already traverse hundreds of complete libration cycles, making them well suited to identifying the stability structure relevant to the mass-limit contours studied here. To assess how this baseline affects our derived mass limits, we repeated the MEGNO analysis for four representative systems (WASP-93 b, Wendelstein-1 b, WASP-20 b, and HAT-P-4 b) up to $10^6$ orbits. These representative validation cases for the present mass-limit workflow show that, although the longer integrations reclassify an additional 15--22\% of marginal edge grid points as unstable, the resulting Trojan mass upper limits change by at most 1\%, because the affected points lie near the outer boundary of the tadpole region. Extending the integrations to $10^6$ orbits would therefore increase the computational cost by a factor of $\sim 100$ while providing only negligible improvement for the mass-limit contours relevant to this study. We therefore adopt $10^4$ orbits as a computationally efficient baseline for the present compatibility analysis, while noting that this comparison is not intended as a general statement about the long-term stability of all 260 systems.

\begin{figure}
  \centering
  \includegraphics[width=0.45\textwidth]{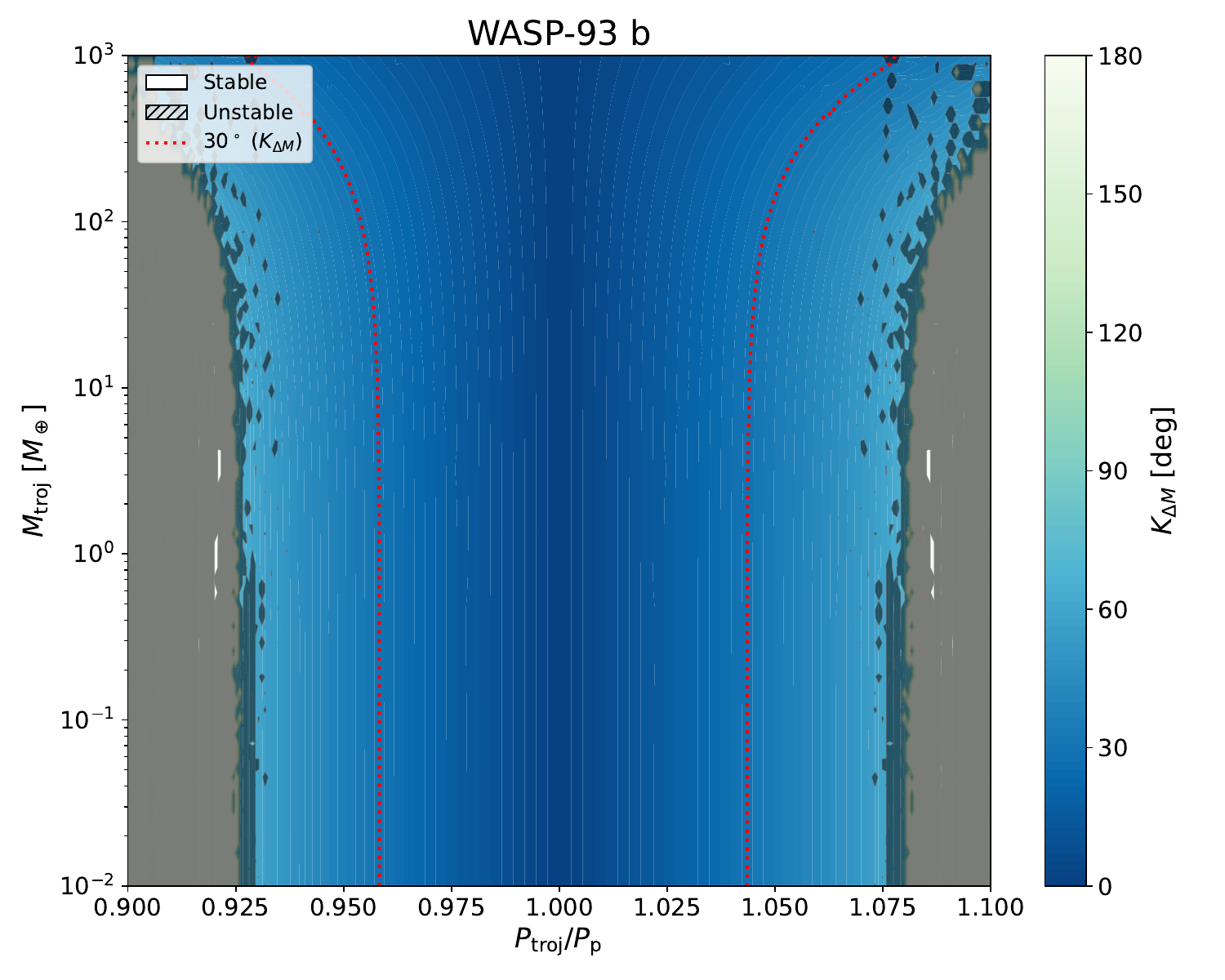}
  \caption{Stability and dynamics heatmap for WASP-93 b. The color scale and contours represent the libration amplitude corresponding to the initial period ratio and Trojan mass. The red dashed line marks the $30^\circ$ libration contour, shown here as a practical reference line within the tadpole zone. Our conservative upper mass limits are strictly derived from the subset of configurations that remain dynamically stable according to the MEGNO mask.
  }
  \label{fig:stability_heatmap}
\end{figure}

We note that for some systems (particularly those with massive planets), stable configurations could not be established for libration amplitudes as large as $30^\circ$, meaning these specific configurations are intrinsically unstable and are therefore automatically excluded from the derived mass limit statistics at these high libration amplitudes. This is consistent with the restricted three-body problem, where the size of the stable region around the Lagrange points decreases as the mass ratio increases \citep{murray2000-solar}. Our final resulting upper mass limits thus specifically denote the boundaries beyond which TTVs are observationally incompatible, considering only those companion configurations that remain strictly dynamically stable. Figure \ref{fig:stability_heatmap} illustrates this approach for the representative system WASP-93 b. The MEGNO mask defines the dynamically stable region, while the $30^\circ$ contour serves as a practical reference line within the tadpole zone used in our discussion of conservative mass limits.

\subsection{Population results}
\begin{figure*}
  \centering
  \includegraphics[width=\textwidth]{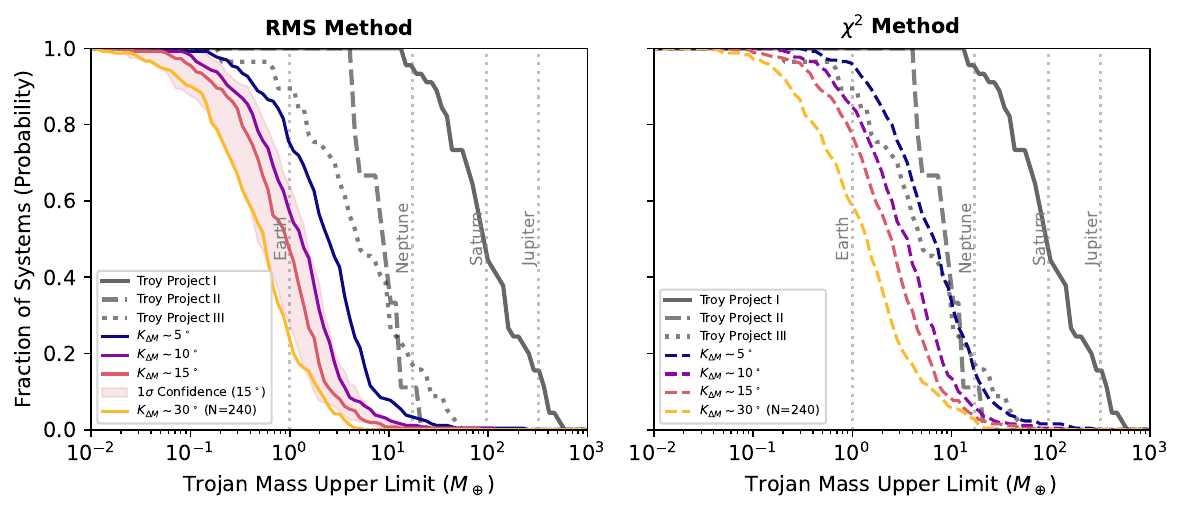}
  \caption{Cumulative probability distribution of the upper mass limit for potential exotrojans in 260 hot Jupiter systems.
    The left panel shows the results from the RMS method, compared with Troy project results in red.
    The right panel displays the results from our $\chi^2$ analysis, offering a more conservative but statistically robust upper limit.
    Reference masses for Neptune, Saturn, and Jupiter are shown.
    Note that systems with unstable orbital configurations at the considered libration amplitudes (e.g., typically around $30^\circ$ for massive planets) are excluded from the statistics. The shaded region represents the $1\sigma$ uncertainty for a libration amplitude of $15^\circ$.}
  \label{fig:cumulative}
\end{figure*}

Our analysis of 260 hot Jupiter systems using CMAT indicates that the upper mass limits for exotrojan companions are strongly dependent on the libration amplitude. For a typical libration amplitude of $15^\circ$ (close to the mean libration amplitude of Jupiter's Trojans, $\sim14^\circ$; \citealt{milani1993-trojan, murray2000-solar}), we find that, with the exception of the two massive super-Jupiters (HATS-70~b and HATS-41~b, both with $M_p \approx 10-13~M_{\mathrm{J}}$), all other studied systems have upper mass limits for Trojans strictly below 10~M$_\oplus$ (with a median limit of 1~M$_\oplus$). The higher mass limits for these super-Jupiters are primarily due to the extreme masses of the primary planets, which dampen the TTV effect of any Trojan. This represents a significant advancement in ruling out massive exotrojans in hot Jupiter systems.
We plot the cumulative probability distribution of the upper mass limit of potential exotrojans in Fig.~\ref{fig:cumulative}. The left panel shows the results based on the RMS method, while the right panel displays the results from our $\chi^2$ analysis. As demonstrated in the right panel of Fig.~\ref{fig:cumulative}, while the $\chi^2$-based limits are more conservative than the RMS-based limits (left panel), they provide a statistically robust exclusion zone. Specifically, based on the RMS method, $50\%$ of the hot Jupiter systems in our study cannot support an exotrojan with a mass exceeding 1~M$_\oplus$, and $80\%$ cannot support one with 2~M$_\oplus$. For the more conservative $\chi^2$ analysis, limits are slightly higher, with $50\%$ and $80\%$ of systems excluding companions more massive than 3~M$_\oplus$ and 6~M$_\oplus$, respectively. Additionally, as discussed in Sect.~\ref{sec:result}, the use of catalog masses (which overestimate $M_{\mathrm{p}}$) makes our derived limits conservative; using the true planet masses would shift the CDF curves in Fig.~\ref{fig:cumulative} to the left.

\subsection{Correlations with planetary parameters}
\begin{figure*}
  \sidecaption
  \includegraphics[width=0.68\textwidth]{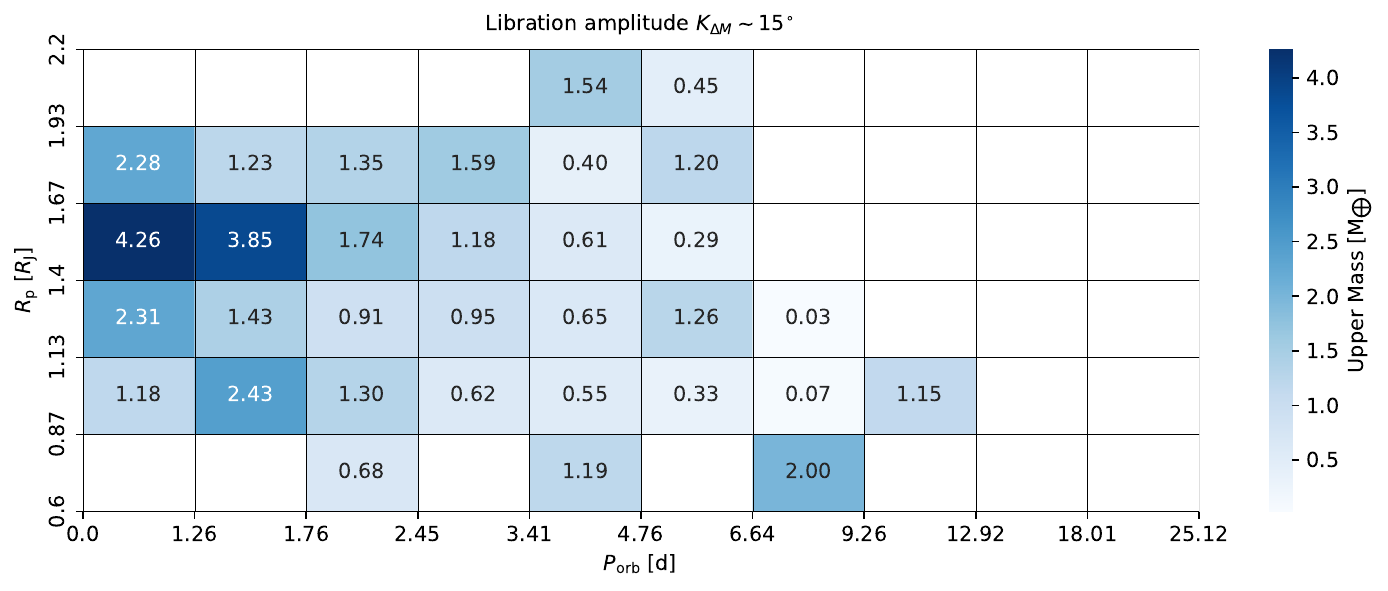}
  \caption{Median upper mass limits of potential exotrojans in 260 hot Jupiter systems, binned by orbital period (1–30~days) and planetary radius (0.5–2.2~$\mathrm{R_{Jup}}$).
    Colors represent the median mass limit in each bin.
  }
  \label{fig:rms}
\end{figure*}

We also sought to identify correlations between the upper mass limits of Trojan companions and the hot Jupiters' parameters shown in Fig.~\ref{fig:rms}. The orbital period ($P$) and radius ($R_{\mathrm{p}}$) of the planets are categorized into 15 $\times$ 6 bins, using a logarithmic scale for the orbital period. The color scheme represents the median upper mass limit in Earth mass for the exotrojans.
We computed Pearson correlation coefficients between the upper mass limits of exotrojan companions and the radii and orbital periods of hot Jupiters. The analysis did not reveal any significant correlations within the current dataset. This may suggest that the current precision of TTV measurements is primarily limited by observational noise rather than being sensitive to subtle physical correlations with planetary parameters.
However, the absence of hot Jupiters in the top right corner bins of Fig.~\ref{fig:rms} suggests that the inclusion of future data in these bins could potentially impact the correlation outcomes.

\section{Discussion}
\label{sec:discussion}
\subsection{Comparison with previous works}
In this section, we compare the findings of this work with those of the Troy Project \citep{lillo-box2018-troy, lillo-box2018-troya, balsalobre-ruza2024-project}. Our study focuses on co-orbital objects associated with hot Jupiters, whereas Troy Project I and II \citep{lillo-box2018-troya} concentrate on exoplanets with short orbital periods ($P<5$ days), and Troy Project III \citep{balsalobre-ruza2024-project} examines systems orbiting low-mass stars.

As illustrated in Fig.~\ref{fig:cumulative}, our results follow a trend similar to previous studies. However, it is important to note the methodological differences: while the TROY project models data directly with exotrojan models to perform a rigorous dynamical characterization, our study presents a compatibility analysis assessing whether TTVs are consistent with a hypothetical Trojan. Consequently, although our sample size is considerably larger (five times that of Troy project I, ten times that of Troy project II, and twice that of Troy project III), our analysis provides a constrained compatibility check rather than the full dynamical modeling of the TROY project. This larger sample size allows for a broad statistical overview, complementing the in-depth targeted studies.

\begin{figure*}
  \sidecaption
  \includegraphics[width=0.68\textwidth]{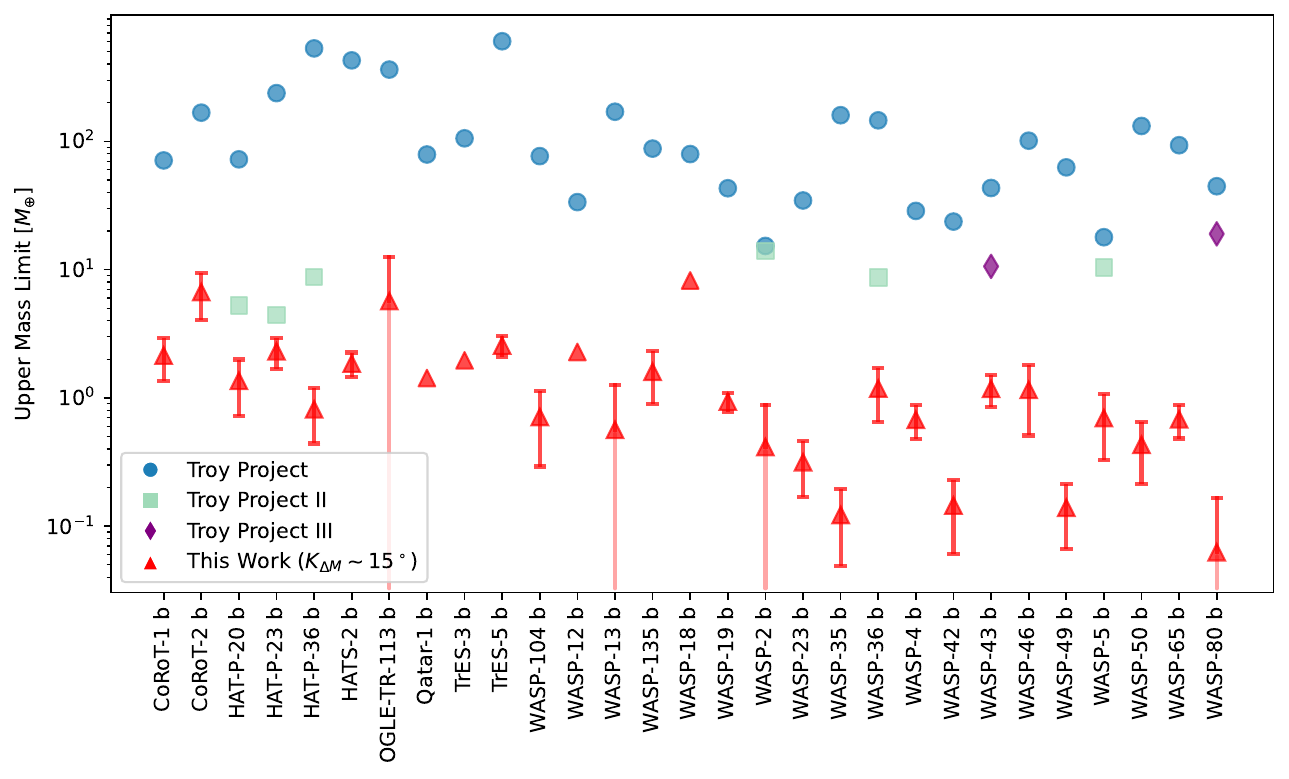}
  \caption{
    The upper mass limits of planets from the Troy project I, II, III, and this work (ExoEcho).
    The x-axis displays planet names, and the y-axis presents mass limits on a logarithmic scale.
    Colors and markers distinguish each study. Red triangles represent our results assuming a libration amplitude of $\sim 15^\circ$. Error bars indicate the $1\sigma$ uncertainty but are omitted when smaller than 15\% of the mass limit for visual clarity. For systems where the lower $1\sigma$ boundary is unconstrained (i.e., consistent with zero mass), we represent the limit using a vertical line extending to the bottom of the axis.
  }
  \label{fig:comparison}
\end{figure*}

Specifically, 29 hot Jupiter systems in our sample have also been studied in the Troy Project. As illustrated in Fig.~\ref{fig:comparison}, our results generally provide improved upper mass limits for exotrojan companions in these systems. This improvement is likely attributable to the high photometric precision and denser observational cadence of TESS, which lead to more precise mid-transit time measurements compared to the ground-based data used in parts of the Troy project.

Furthermore, we note that for the comparison in Fig.~\ref{fig:comparison}, we adopted a typical libration amplitude of $15^\circ$ (consistent with the mean libration of Jupiter Trojans; \citealt{milani1993-trojan, murray2000-solar}). These sensitivity limits are therefore biased towards this specific libration amplitude; as demonstrated in Sect.~\ref{sec:result}, a smaller libration amplitude would result in weaker mass constraints. This dependency on the dynamical configuration is a key limitation of our method compared to full dynamical fits.

In Fig.~\ref{fig:comparison}, we present the upper mass limits derived from our TTV analysis. The $1\sigma$ uncertainties correspond to the intersection of observed TTV RMS bounds with our simulation grids. For systems where the lower bound of the upper mass limit is unconstrained (consistent with zero mass at the $1\sigma$ level), we indicate this using vertical lines extending to the bottom of the axis. While most analyses assume a libration amplitude of $15^\circ$, we adopted $14.5^\circ$ for HAT-P-20 b to ensure orbital stability. For improved visual clarity, error bars smaller than 15\% of the mass limit have been omitted.

\subsection{Degeneracy between mass and libration amplitude}
A key constraint of the TTV method is the degeneracy between the exotrojan mass and the libration amplitude (Eq.~\ref{eq:ttv-amplitude}). We addressed this by scanning a range of period ratios to simulate various libration amplitudes. Consequently, the mass limits presented in this work are explicitly defined as a function of the libration amplitude. While a smaller true libration amplitude would indeed allow for a more massive companion to remain undetected, our results provide a comprehensive map of these compatibility limits across the dynamical parameter space.

\subsection{Impact of eccentricity}
\begin{figure}
  \centering
  \includegraphics[width=0.45\textwidth]{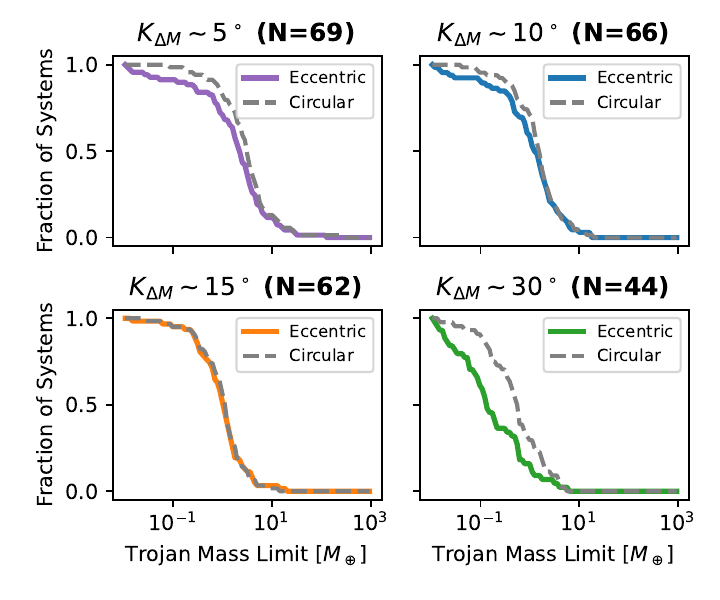}
  \caption{Cumulative probability distribution of the upper mass limit for a subset of systems with eccentricity $e>0.05$. The gray dotted line represents the results assuming circular orbits, while the colored solid line shows the results using the measured eccentricities. The inclusion of eccentricity generally tightens the mass constraints. Similar to Fig.~\ref{fig:cumulative}, systems that are dynamically unstable at the tested libration amplitudes are excluded.
  }
  \label{fig:eccentric_subset}
\end{figure}

We acknowledge that assuming circular orbits is a simplification, as hot Jupiters often possess non-zero eccentricities. While our bulk analysis maintains this assumption due to the typically low eccentricities of these planets, we recognize that eccentricity can introduce systematic effects. To quantify this, we performed additional simulations using measured eccentricities for a subset of 82 systems with $e>0.05$ (Fig.~\ref{fig:eccentric_subset}). We find that including eccentricity generally tightens the derived mass limits while preserving the overall trend. Interestingly, the impact of eccentricity on TTV amplitude is non-linear with respect to libration amplitude, with the effect being minimized at $\sim15^{\circ}$. This minimum coincides remarkably well with the mean libration amplitude of Jupiter's Trojans ($\sim14^{\circ}$; \citealt{milani1993-trojan}), although the physical significance of this agreement remains unclear. Finally, we note that for systems with moderate eccentricities, the stable point for a 'quiet' configuration (small libration) shifts away from the exact $60^\circ$ separation, a dynamical feature that must be accounted for in accurate definitions of libration amplitude \citep{namouni1999-secular}.

\subsection{$\chi^2$ analysis}
Additionally, we incorporated a $\chi^2$ analysis to complement the RMS method. While the RMS approach provides stringent constraints by globally aggregating TTV scatter, it treats all data points equally irrespective of their individual observational uncertainties. In contrast, the $\chi^2$ analysis explicitly weights deviations by their measurement errors, providing a more statistically robust assessment of compatibility. This method prevents data points with high scatter and large uncertainties---which are common in ground-based follow-up observations or fainter TESS targets---from artificially inflating the apparent TTV signal.

Our results show that the $\chi^2$-based mass limits are generally more conservative (i.e., higher) than those derived from the RMS method. For instance, while the RMS method excludes companions more massive than 1 $M_\oplus$ for 50\% of the systems, the $\chi^2$ analysis places this threshold at 3 $M_\oplus$. This difference underscores that while the RMS metric serves as a useful proxy for overall variability, the $\chi^2$ statistic provides a more robust exclusion criterion by incorporating data quality. Despite these quantitative differences, the consistency between the overall trends of the RMS and $\chi^2$ results strengthens the validity of our general conclusions concerning the scarcity of massive exotrojans in these systems.

\subsection{Future prospects}
Future work will expand this study by including less massive exoplanets in our sample.
Since the TTV amplitude is related to the mass ratio between the exotrojan and the target planet (Eq.~\ref{eq:ttv-amplitude}), we expect to have much stronger constraints on their exotrojans for less massive target planets.

\section{Conclusion}
\label{sec:conclusion}

As part of the ExoEcho project, this study presents a comprehensive analysis of the compatibility limits of exotrojans in 260 hot Jupiter systems using the TTV technique with the latest data from the TESS mission.
Combining TTV data from the target planet with N-body simulations, we have been able to put constraints on the mass of exotrojans in 260 hot Jupiter systems, mostly below $10~M_{\mathrm{\oplus}}$, while assuming a libration amplitude of $15{\si{\degree}}$.
Thanks to the all-sky survey capability of TESS, our study has one of the largest sample sizes thus far among studies focusing on exotrojan detection.
Our constraints on exotrojan masses in these systems, reinforced by $\chi^2$ analysis, provide new statistical limits on co-orbital dynamics in the hot Jupiter regime.

The libration of exotrojans around the Lagrangian points induces TTVs of the hot Jupiters, which are typically non-periodic and thus more challenging to detect in photometric observations \citep{janson2013-systematic}.
Additionally, an inclined orbit can cause these exotrojans to avoid detection during transits, and their small masses make them difficult to identify through radial velocity methods.
Continued transit observations from TESS and future space missions like PLATO \citep{rauer2024-plato} and ET \citep{ge2024-searcha} will allow for more sensitive mass limits, further constraining theories of exotrojan formation.

Finally, it is crucial to reiterate the limitations of the TTV technique in this context. Our derived upper mass limits are sensitive to the dynamical state of the potential companion; specifically, our method is blind to exotrojans with negligible libration amplitudes, as they do not generate detectable TTV signatures. Consequently, the absence of a TTV signal excludes only those massive companions that are undergoing significant libration.

\section*{Data availability}
The full machine-readable version of Table~\ref{tab:ttv-data}, containing the measured mid-transit times for all 260 systems, is available at the CDS.

\begin{acknowledgements}
  We express our gratitude to the TESS mission for its outstanding contribution to exoplanet science. The precise transit-time measurements from TESS were instrumental in our study of TTVs in hot Jupiters. We thank NSFC grants 12073092 and 12103098, the Natural Science Foundation of Hainan Province (Grant Nos. 125RC632 and 424QN217), the National Key R$\&$D Program of China (2020YFC2201400), the science research grants from the China Manned Space Project (No. CMS-CSST2021-B09), and the Earth 2.0 Space Mission program from SHAO for their financial support. Additionally, we appreciate the valuable assistance of the NASA Exoplanet Archive, which granted us access to a vast amount of observational data and resources that greatly aided our research efforts. We also thank open-source software developers for their contributions to the scientific community.
\end{acknowledgements}

\bibliographystyle{aa}
\bibliography{main}
\end{document}